\documentclass[%
 reprint,
 amsmath,amssymb,
 aps,
]{revtex4-2}

\usepackage{graphicx}
\usepackage{dcolumn}
\usepackage{bm}

\usepackage{subfig}

\begin{document}

\preprint{XXXXXXXXXX}

\title{Rumour Spreading In Community Based Networks}%

\author{Zhaoxi Cui}
\affiliation{%
 Computing Science and Mathematics, University of Stirling\\
 Stirling, FK9 4LA. United Kingdom
}%

\author{Anthony O`Hare}
\affiliation{%
 Computing Science and Mathematics, University of Stirling\\
 Stirling, FK9 4LA. United Kingdom
}%

\date{\today}

\begin{abstract}
Many real-world networks have the characteristic that they are comprised of distinct groups or communities whose members contain many links within the community but with fewer connections to others. 
It is important to accurately model these types of networks to correctly predict the outcome of important spreading processes such as disease transmission, or the flow of information etc.
Our motivating example is a network of traders within several investment institutions such as hedge funds. We assume an idealised scenario where traders within the same institution have many contacts and can share information quickly and easily but have fewer contacts to traders in other institutions, relying on personal networks, allowing for information to flow easily within a community and less-so between communities.
In this paper we investigate a particular spreading process, the spread of a rumour, on a community based network that is characterised by two parameters; the within-group connectivity, and the between-group connectivity. We show that such networks have different characteristics to small-world or random networks that are often used to model the types of systems and that the network topology has a small but not insignificant  effect on the spread of rumours on the network.
\end{abstract}

\keywords{
rumour \sep spread \sep community \sep networks \sep financial markets
\PACS 02.10.Ox \sep 89.65.Gh
\MSC 91D30 \sep 91G15 \sep 91G45}
                              
\maketitle


\section{Introduction}
\label{sec:introduction}

One of the earliest models to describe the spread of rumours used a probabilistic approach to describe the transitioning of individuals between different states based on their exposure to a rumour \cite{Rapoport1948, DaleyKendall1965} in a similar way epidemic models are used to study the spread of disease in a population \cite{KermackMcKendrick1927}. 
Daley and Kendall \cite{DaleyKendall1965} formulated a model to simulate the spread of rumour, by conceptualising a closed, homogeneously mixing population  and dividing it into three mutually exclusive and exhaustive categories; Ignorants (\(I\)), individuals who have not heard the rumour, Spreaders (\(S\)), individuals who have heard the rumour and are actively spreading it, and Stiflers (\(R\)), individuals who have heard the rumour but have stopped spreading it, either due to disbelief, lack of interest, or because the rumour has reached a saturation point where it is no longer deemed `news'.
The DK model was formulated as a system of differential equations to describe the transitions between these states, drawing parallels to the SIR (Susceptible-Infected-Removed) model in epidemiology \cite{KermackMcKendrick1927}.

While the Daley-Kendall (DK) model provides a foundational understanding of rumour dynamics, subsequent research has introduced several variations and extensions to better capture the complexities of real-world information spread. These modified models incorporate additional parameters and mechanisms, allowing for more nuanced representations of how rumours propagate through diverse social networks. 

The Maki-Thompson model \cite{MakiThompson1973} simplifies the Daley-Kendall model by focussing on direct contacts rather than pairwise interactions and only the initiating spreader becomes a stifler when meeting another spreader. The Maki-Thompson model was one of the first to incorporate stochastic elements to account for the inherent randomness in rumour spreading; using a Markov chain framework to describe the transitions between states, providing a more realistic representation of rumour dynamics.

Despite similarities with  epidemiological models, the distinct recovery rule causes fundamental differences in the rumour-free equilibrium. In contrast to the self-decay process in the disease spreading model, the stifling mechanism in the rumour spreading process highlights a contact-based transition; the spreaders  become stifled only as a result of encountering those who are already aware of the rumour whereas in an epidemiology model,  infected individuals recover such that the removal is a linear function of the overall number of infected. 

More recently, network structure has been incorporated into rumour propagation models \cite{Zanette2001}. Once such model considered the small-world properties of social networks \cite{Zanette2001} demonstrating that the structure of the network significantly affects the spread of rumours, with small-world networks facilitating faster dissemination compared to random networks.

The introduction of scale-free networks by Barab\'{a}si and Albert in 1999 \cite{Barabasi1999} further advanced the field . These networks, characterised by a few highly connected nodes (hubs), were found to be particularly effective in spreading information \cite{Wan2017,Brisson2018}. Researchers adapted the DK model to scale-free networks, revealing that rumours spread more efficiently in such networks due to the presence of these hubs.

In recent years, extensions of the basic models have been proposed to capture more complex behaviours. For instance, the ISS (Ignorant-Spreader-Stifler) model \cite{Sun2021, Yu2024, wang2021spreading} was extended to include a verifier state, resulting in the ISSV model \cite{al2015qualitative}. These models account for individuals who actively verify the information before spreading it, reflecting the impact of fact-checking in modern information dissemination.

In these rumour spreading models, a spreading threshold has been found to exist \cite{liu2024dynamic, tong2022dynamic}, similar to the epidemic threshold in disease models \cite{KermackMcKendrick1927, keeling2008modeling, pastor2001epidemic}.
This threshold determines whether a rumour will die out quickly or spread widely through the population. Models have shown that network structure, initial conditions, and individual behaviours all influence this threshold.

The study by Nekovee et al. \cite{nekovee_theory_2007} introduces a comprehensive stochastic model for analysing the dynamics of rumour spreading across complex social networks, with an emphasis on both homogeneous and heterogeneous network structures, including random graphs and scale-free networks \cite{nekovee_theory_2007}. The authors build on the established frameworks of the Maki–Thompson (MK) rumour model and the Susceptible-Infected-Removed (SIR) epidemic model to propose a unified approach that effectively captures the dynamics of information dissemination in complex social contexts. 

A critical contribution of this work is the identification of a threshold behaviour in the context of rumour propagation. The authors demonstrate that in homogeneous and random networks, the model exhibits a well-defined critical threshold for the spreading rate, below which a rumour fails to propagate throughout the system. This threshold is reminiscent of the epidemic threshold found in SIR models, suggesting a similar underlying mechanism in the dynamics of contagion-like processes in these networks. Conversely, in scale-free networks, which are characterised by their heavy-tailed degree distribution, the critical threshold vanishes as the network size becomes infinitely large. This phenomenon may be applicable in modelling the spread of rumours within social media networks where there exists a disproportionately large influence of highly connected nodes, or hubs.

Furthermore, the study by Nekovee et al. \cite{nekovee_theory_2007} investigates the influence of network topology on both the speed and extent of rumour propagation. The authors report that the initial rate of rumour dissemination is significantly higher in scale-free networks compared to random graphs. This is attributed to the presence of hubs that serve as efficient conduits for information, rapidly spreading the rumour to a large number of nodes. Interestingly, the introduction of assortative degree correlations—a feature where highly connected nodes preferentially attach to similarly connected nodes—further accelerates the initial spreading rate. However, these assortative correlations have a dual effect; while they enhance the early spread, they can also limit the ultimate reach of the rumour. Once hubs transition from being spreaders to stiflers, they effectively hinder further dissemination, highlighting a complex interplay between network structure and spreading dynamics.

In terms of the model's steady-state behaviour, the authors present both analytical and numerical solutions that elucidate the critical behaviour and equilibrium dynamics across different network types. Their findings reveal that the impact of assortative correlations on the final size of the rumour is contingent on the spreading rate. Specifically, in assortatively correlated networks, the final fraction of nodes that receive the rumour is initially smaller compared to uncorrelated networks at lower spreading rates. However, as the spreading rate increases, the assortative network demonstrates a higher final rumour size, illustrating the nuanced role of assortative mixing in rumour propagation. This study concludes by emphasising that scale-free networks are inherently susceptible to rumour spreading, similar to their susceptibility to epidemics. This susceptibility is largely due to the role of hubs, which not only facilitate rapid dissemination but also significantly influence the overall dynamics of the spread. The authors suggest that future research should explore the effects of dynamic network structures, such as those observed in internet-based social networks like chatrooms, where the time scales of network topology changes may align with those of the spreading process. Such dynamic environments present unique challenges and opportunities for understanding the full complexity of rumour propagation in real-world settings.

\section{Methods}
\label{sec:methods}

Our motivation in this manuscript is to understand how information, for example a rumour, spreads through a network of individuals employed in distinct communities where information can flow easily within a community and less so between them. For example, financial traders can share information with colleagues within their department but rely on fewer personal contacts with traders in other institutions.
In this idealised scenario, we envisage two types of connections; within-group connections (for those that are within the same institution and where there are low barriers to the flow of information and so it can spread at a relatively fast rate) and between-group connections (where there are fewer connections between individuals in different institutions/groups and so information flows at a slower rate).

To capture the structural features of real-world social communities, we construct a network of $N$ nodes, denoting trading agents, and $E$ edges representing the pair-wise contacts between them. 

We divide the network into $N_g$ communities and denote the connectivity of agents within the same community as $w_g$ and the connectivity of agents between separate communities as $b_g$. 
Both $w_g$ and $b_g$ are parameterised between 0 and 1 and we interpret them as the probability that individuals or nodes within a community are connected to other nodes within the same group or community and between nodes in different groups or communities. 
Thus the communities are built up from $N_g$ random-networks which are randomly connected with each other.

We create the community-based network in the following manner:
We assign the $N$ nodes in the network to one of the $N_g$ communities. 
For each pair of nodes in the network, if they are in the same community/group we connect them with a probability $w_g$ and if they are in a different group we connect them with a probability $b_g$. 
We naturally assume that $w_g > b_g$ in our idealised scenario but do not enforce this in our network building algorithm or in our analysis.

To characterise this network, we measure the level of clustering in terms of $w_g$ and $b_g$ using two quantitative measures clustering coefficient and modularity, which are then compared against an Erdos-Renyi random network and a Watts-Strogatz small-world network.

In this study, we employ the clustering coefficient as a key metric to quantify the level of local clustering within networks. The clustering coefficient measures the tendency of nodes in a network to form tightly connected groups, or cliques. Mathematically, the local clustering coefficient for a node \( i \) is defined as:
\[
C_i = \frac{2T_i}{k_i(k_i-1)},
\]
where  \( T_i \) is the number of triangles (closed triplets, i.e.  groups of three nodes that are all connected) involving node \( i \),  \( k_i \) is the degree of node \( i \), and \( k_i(k_i-1)/2 \) is the maximum possible number of triangles for a node with degree \( k_i \).

The mean clustering coefficient, used throughout this work, is the average of \( C_i \) over all nodes in the network:
\[
C = \frac{1}{N} \sum_{i=1}^{N} C_i,
\]
where \( N \) is the total number of nodes. A higher clustering coefficient indicates a greater prevalence of tightly-knit groups within the network.

The clustering coefficient is particularly relevant for understanding how network structure fosters local interactions, such as those within communities, and it provides a direct measure of the community structure of a network, complementing global metrics such as modularity.
We use the clustering coefficient to evaluate the extent to which local clustering persists under varying intra-group (\(w_g\)) and inter-group (\(b_g\)) connection probabilities. In this work:
\begin{itemize}
    \item[] For \textbf{Community Based Networks} that are the subject of this research, we expect high clustering coefficients due to their explicit design, which emphasises dense intra-group connections.
    \item[] For \textbf{Random Networks}, clustering coefficients are anticipated to remain low, reflecting the randomness of connections and lack of inherent clustering.
    \item[] For \textbf{Small-World Networks}, we expect intermediate values, as these networks balance local clustering with long-range connections.
\end{itemize}

Each node's expected number of edges (its centrality) can be expressed as: 
\begin{equation}
   C = \frac{w_g N}{N_g} + b_g\left(N-\frac{N}{N_g}\right)
\end{equation}
where $w_g \frac{N}{N_g}$ is to the number of contacts made within its cluster while  $b_g (N-\frac{N}{N_g})$ is the number of connections to outsiders. 

Normalization gives the expected degree centrality (the number of links incident upon a node) for each node in community based network as
\begin{equation}
   \langle C \rangle= \frac{w_g}{N_g} + b_g\left(1-\frac{1}{N_g}\right)
   \label{eqn:clusteringCoeff}
\end{equation}

To understand the clustering properties in the community based network, let us first look at the boundaries of the $w_g, b_g$ domain, by setting one of the parameters to 0 and vary the other from [0,1]. 

\begin{description}
\item[$\mathbf{b_g=0, w_g=\left[0,1\right]}$] At $(w_g, b_g) =(0,0)$ no nodes in the network are connected. Increasing the within group connectivity leads to a number of disjoint sub-networks each of which become complete at $w_g=1$. At this point we have $N_g$ complete sub networks and so the clustering coefficient has its maximum.
\item[$\mathbf{w_g=0, b_g=\left[0,1\right]}$] Here we increase the number of between group connections while keeping the groups themselves disconnected. At $b_g=1$, each node is connected to all the nodes outside of its community but to none within their community at which point it is maximally connected and has a maximum clustering coefficient. Closed triangles occur when nodes are selected from different communities, and open triangles occur 2 of the nodes in the triplet are selected from the same community. 
\item[$\mathbf{b_g=1, w_g=\left[0,1\right]}$] Every node in a community is completely connected to nodes outside the community but has zero connections within the community when $w_g=0$. By increasing the within-group connections, $w_g$, the number of closed triplets increases while open ones are reduced to $0$ at $w_g=b_g=1$ the network becomes complete and becomes one large cluster.
\item[$\mathbf{w_g=1, b_g=\left[0,1\right]}$] The clustering coefficient reaches the maximum at $b_g=0, w_g=1$ where the network consists of disjoint complete networks, adding connections to any pair of clusters would create open triplets decreasing the clustering coefficient.
\end{description}

\subsection{Comparison to Random Networks}

An Erdos-Renyi random network \cite{ErdosRenyi1959} has two parameters: the number of nodes, $N$, and the probability of an edge occurring between any two nodes, $p$. As $p\to1$ the network becomes complete, i.e. all nodes are connected to all others, at which point the network will have its maximum clustering coefficient. The probability $p$ can be expressed in terms of $w_g, b_g$ as 

\begin{equation}
   p = \frac{w_g}{N_g} + b_g \left(1-\frac{1}{N_g}\right)
\end{equation}
where $N_g$ is the number of communities in the network. This allows us to directly compare the clustering of community based networks with random ones.

\subsection{Comparison to Small World Networks}
Random networks do not, typically, have local clustering since the probability of two nodes being connected is constant, random, and independent. Another important network that is used to model information spread is the small-world network. The Watts-Strogatz \cite{WattsStrogatz1998} algorithm for creating small world networks is characterised by 2 parameters; $K$, the number of nearest-neighbours a node is connected to, and $p$, the probability that a neighbouring node is re-wired to another node.  

Let us now express these in terms of $w_g$ and $b_g$. The probability that 2 nodes are within the same community is approximately $K(1-p)$ and the probability that 2 nodes are in different communities is $Kp$.
Setting the number of connections within a group and between groups for the community based and small world networks equal,  we have
\begin{gather}
    w_g \frac{N}{N_g} = K(1-p) \nonumber \\
    b_g\left(N-\frac{N}{nN_g}\right) = Kp
\end{gather}
with some rearranging gives
\begin{align}
        K &= N\left(w_g\frac{1}{N_g} + b_g\left(1-\frac{1}{N_g}\right)\right) \nonumber \\
        p &= \frac{b_g(N_g-1)}{w_g+b_g(N_g-1)}
    \label{eqn:smallwrld}
\end{align}

\subsection{Comparison of Network Structure}

\begin{figure}[ht!]
\includegraphics[width=0.8\textwidth]{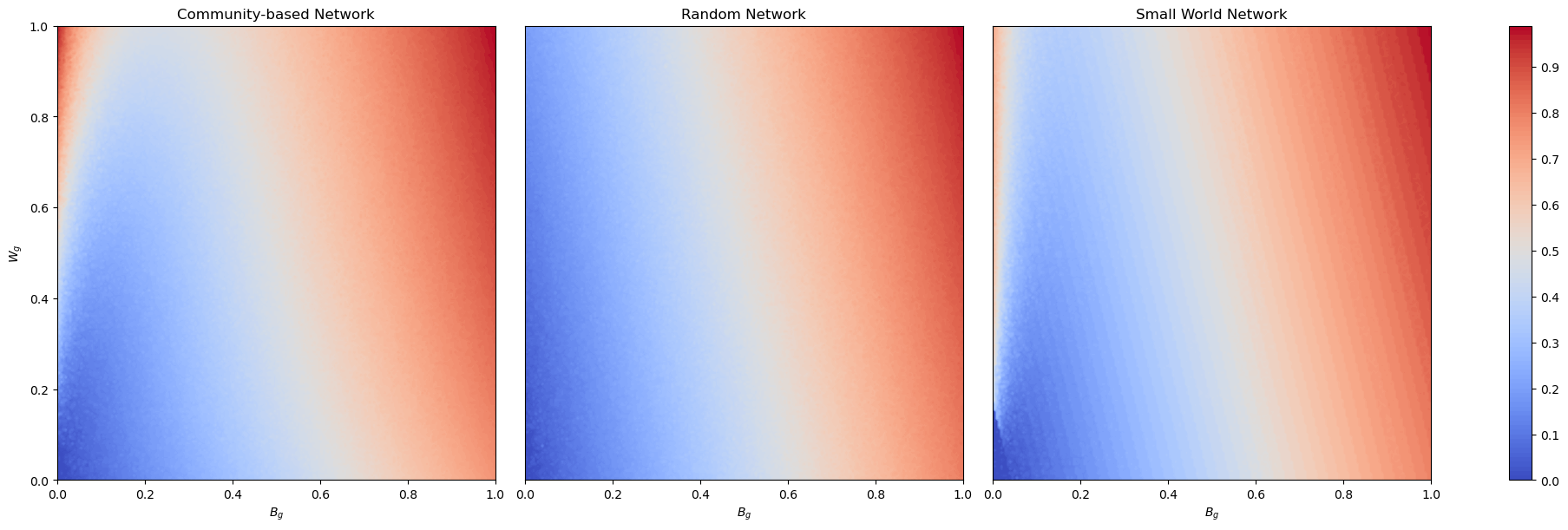}
\caption{Contour plots of the mean clustering coefficient for community-based (left), random (centre), and small world (right) networks. Each network consists of $N$=1000 nodes with $N_c$=40 communities. For the small world network $p=\frac{b_g(N_c-1)}{w_g+b_g(N_c-1)}$, $K=round(N (\frac{w_g}{N_c} + b_g(1 - \frac{1}{N_c}))$, and for a random network $ p = \frac{w_g}{N_g} + b_g\left(1-\frac{1}{N_g}\right)$. }
\label{fig:clust_coeffs} 
\end{figure}

The clustering coefficient quantifies the propensity of nodes to form tightly-knit local groups. Plots of the clustering coefficient for the different networks over a range of \(w_g\) and \(b_g\) show distinct differences at low between-group connection levels, figure \ref{fig:clust_coeffs}.

In community based networks, the clustering coefficient is highest in the region where \(w_g\) is large and \(b_g\) is small as expected, as strong intra-group connectivity fosters dense local clusters within communities, while minimal inter-group connectivity ensures these communities remain well-isolated. As \(b_g\) increases, the introduction of inter-group connections weakens this isolation, leading to a gradual reduction in the clustering coefficient. Similarly, as \(w_g\) decreases, the lack of intra-group connections diminishes the network's ability to sustain strong local clustering. 

\onecolumngrid

\twocolumngrid

For random networks, the clustering coefficient remains consistently low across all combinations of \(w_g\) and \(b_g\). This is a 
because edges are placed randomly without regard for group structure. The lack of deliberate clustering or community features results in a network with minimal local clustering, regardless of the connectivity parameters. This uniformity highlights the random network's inability to exhibit meaningful local clustering.

The small-world network exhibits intermediate clustering levels between those of the community and random networks. When \(w_g\) is large and \(b_g\) is small, the clustering coefficient is relatively high due to the presence of tightly connected local neighbourhoods. However, as \(b_g\) increases, the randomness introduced by the rewiring process dilutes these local connections, resulting in a steady decline in the clustering coefficient. Unlike the random network, the small-world network retains some local clustering even at higher \(b_g\), reflecting its balance between local and global connectivity, but this decays faster than in community based networks.

The \(w_g\)-high, \(b_g\)-low region offers a particularly striking contrast between the three network types. In this region, the community network stands out with a significantly higher clustering coefficient since the network is characterised by local cliques (or communities) that are only loosely connected (if at all) to other cliques. While the small-world network also demonstrates reasonably high clustering in this area, the presence of long-range rewired connections prevents it from matching the clustering levels observed in the community network. 

Modularity, as a measure of community structure, quantifies the extent to which nodes within the same group are densely connected while connections between groups remain sparse.

\begin{figure}[ht!]
\includegraphics[width=0.99\textwidth]{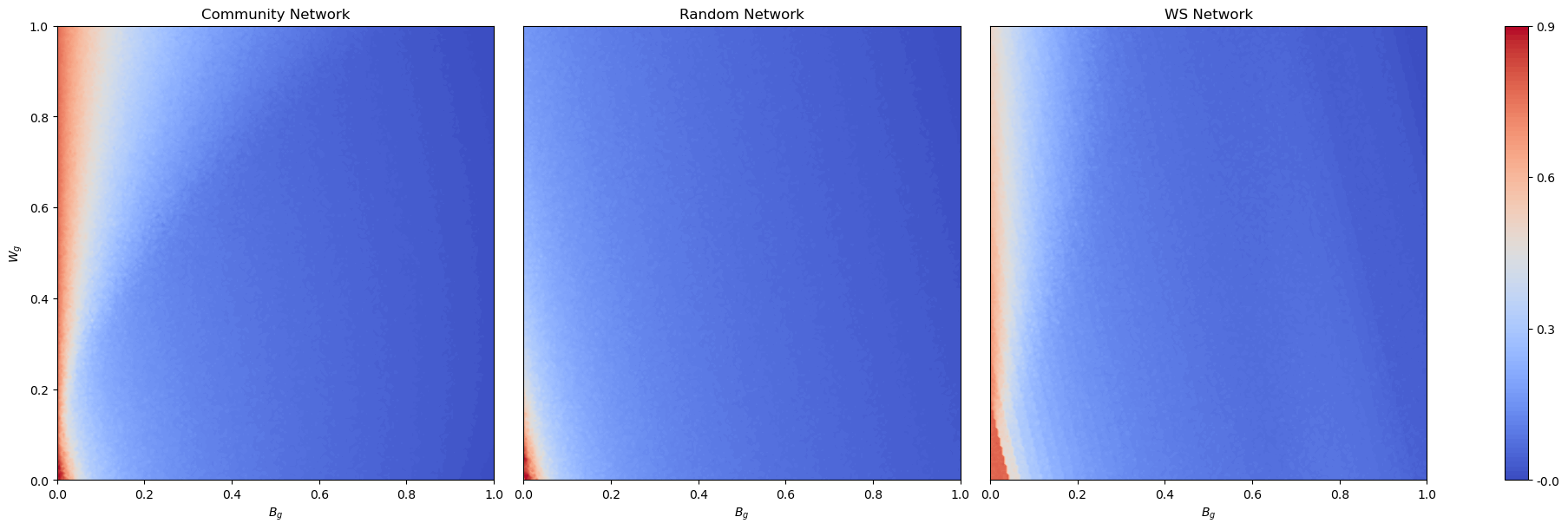}
\caption{Contour plots of mean clustering coefficient for community-based (left), random(centre), and small world (right) networks. Each network consists of $N$=1000 nodes with $N_c$=40 communities.}
\label{fig:modularity} 
\end{figure}

For  \(b_g=0\) there are no inter-group connections in the community networks, while the chance of inter-group connections in both random and small-world networks is low but not impossible, as shown in figure \ref{fig:modularity}. Without inter-group connections, the separation between communities remains intact, resulting in higher modularity, as the network is effectively divided into distinct clusters or isolated nodes 

\onecolumngrid

\twocolumngrid

As \(b_g\) increases, the modularity declines with the erosion of community boundaries due to the increased influence of inter-group connections. 

The random network exhibits uniformly low modularity across all values of \(w_g\) and \(b_g\) since the network lacks any deliberate clustering or structural features that promote community formation. 

The small-world network demonstrates modularity levels that fall between those of the community and random networks. At high \(w_g\) and low \(b_g\), the modularity is high, despite of relatively lower sensitivity to \(w_g\) compared to that of the community network. However, the rewiring probability, \(p\), introduces randomness through long-range connections, which weakens the overall community structure. Consequently, the modularity values for the small-world network are consistently lower than those of the community network, particularly as \(w_g\) increases. 

In the region where both \(w_g\)  and \(b_g\) approach zero, all three network models exhibit relatively high modularity.  Modularity measures the fraction of edges that fall within communities relative to a null model, but in sparsely connected networks, the few existing edges are likely to fall within the same small clusters, inflating modularity scores. 

We also employ \textit{local efficiency} as a complementary metric to assess the robustness and resilience of local connectivity within networks. While the clustering coefficient quantifies the prevalence of tightly connected groups (or cliques) by measuring the likelihood that a node's neighbours are also connected to each other, local efficiency extends this concept by evaluating how efficiently information can propagate within a node’s local neighbourhood, even in the absence of the node itself.
Mathematically, the local efficiency of a node \( i \) is defined as:
\[
E_i = \frac{1}{k_i(k_i-1)} \sum_{j,k \in \mathcal{N}_i} \frac{1}{d_{jk}},
\]
where \( \mathcal{N}_i \) represents the set of neighbours of node \( i \), \( d_{jk} \) is the shortest path distance between neighbors \( j \) and \( k \) within the subgraph excluding node \( i \), and \( k_i(k_i-1) \) normalises the sum by the number of possible connections among the neighbours.

The \textbf{global local efficiency} of the network is then obtained by averaging \( E_i \) over all nodes:
\[
E_{\text{local}} = \frac{1}{N} \sum_{i=1}^{N} E_i.
\]

\onecolumngrid

\begin{figure}[ht!]
\includegraphics[width=0.99\textwidth]{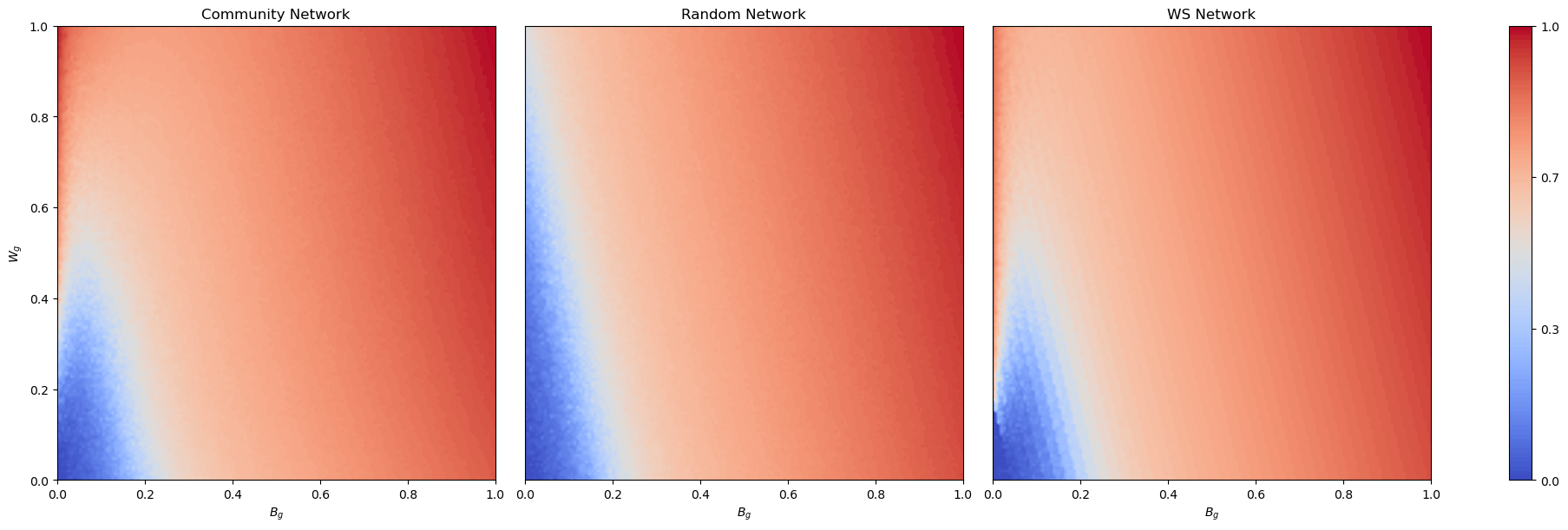}
\caption{
Contour plots of local efficieny for community-based (left), random(centre), and small world (right) networks. Each network consists of $N$=6000 nodes with $N_c$=100 communities. }
\label{fig:local_efficiency} 
\end{figure}

\twocolumngrid

Figure \ref{fig:local_efficiency} presents the local efficiency for the three networks. Similar to the clustering coefficient, the local efficiency sees an overall increasing trend from \( (w_g, b_g) = (0,0) \to (1,1)\). Like the modularity and clustering coefficients, the networks differ slightly when the connections between communities is low. In this region, community based networks have a slight advantage in efficiency over small-world networks.

\section{Information spreading in Networks}

Let us now describe a modification of the Maki-Thompson model \cite{MakiThompson1973} on our networks where the nodes represent agents and edges represent a contact between nodes where information can spread. Agents are assigned a status: \textit{ignorant (I)}, for those who have not heard the rumour, \textit{spreader (S)}, for those that are spreading the rumour, and \textit{stifler (R)}, for those that have heard the rumour and are no longer spreading it. The assumption is that an individual who is ignorant of the rumour will become  a spreader upon contact with another spreader, if two spreaders meet only one (the one initiating the contact) will be come a stifler, and if a spreader meets a stifler the spreader becomes a stifler.
The transitions between the compartments can be summarised as: 
\begin{equation}
      \begin{cases}
      S + I \xrightarrow{\alpha} S + S \\
      S + S \xrightarrow{\theta} S + R\\
      S + R \xrightarrow{\theta}  R + R\\
      \end{cases}
      \label{eqn:rumour_model}
\end{equation}
where $\alpha$ is the spreading rate, the number of contacts between ignorant and spreaders that result in an ignorant becoming a spreader per unit time, 
and  $\theta$, the stifling rate, the number of contacts between spreader and other spreaders or stiflers that result in a spreader becoming a stifler per unit time.

We run several simulations of this process on the different networks already described and compare them using  four representative metrics. 
\begin{description}
\item[Maximum rumour size, $S_\text{max}$,] the maximum number of spreaders as a proportion of the number of individuals in the network.
\item[Time to reach maximum rumour size, $T_\text{Smax}$,]  the [first] time when the number of spreaders is at its maximum. 
\item[Final stifler size, $R^*$,] the number of stiflers at equilibrium as a proportion of the total population, this reflects the ultimate proportion who have heard the rumour.
\item[Time to reach equilibrium state, $T_{eq}$,] the time of stifler reaches the maximum, (when the process reaches equilibrium).
\end{description}

These metrics provide a summary of the dynamics, capturing both the intensity (e.g., peak spreader proportion, final stifler proportion) and timing (e.g., time to peak and equilibrium) of the spreading process.

\section{Results}

We performed Monte Carlo simulations of the model  described in eqn. \ref{eqn:rumour_model} on networks with 6000 nodes, a community size, $N_c=100$, with $\alpha=0.001$ and $\theta=0.001$. We choose $b_g=0.0005$ as its in the region of phase space identified by the clustering coefficient where we expect to see differences in the networks. The parameter $w_g$ is uniformly incremented from $0$ to $1$.

\begin{figure}
  \setlength{\tabcolsep}{0pt}
  \begin{tabular}{cc}
    \includegraphics[width=0.5\textwidth]{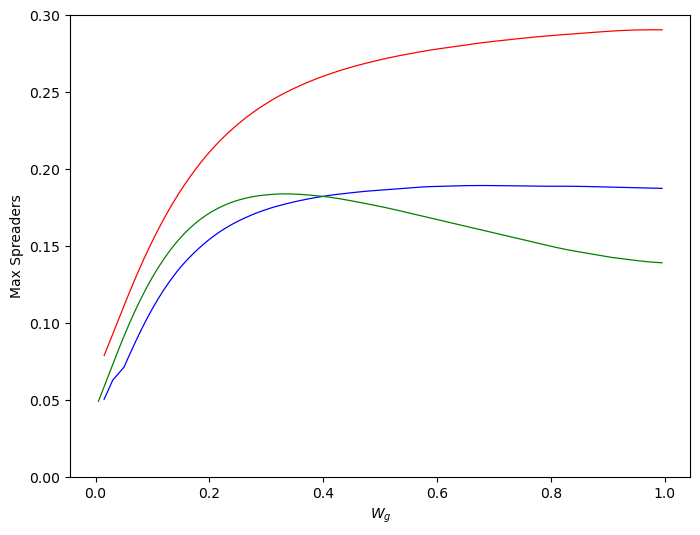} & \includegraphics[width=0.5\textwidth]{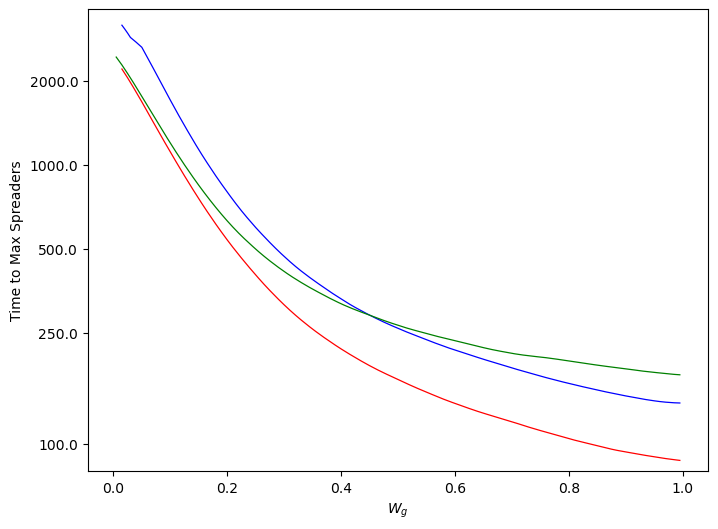} \\
    \includegraphics[width=0.5\textwidth]{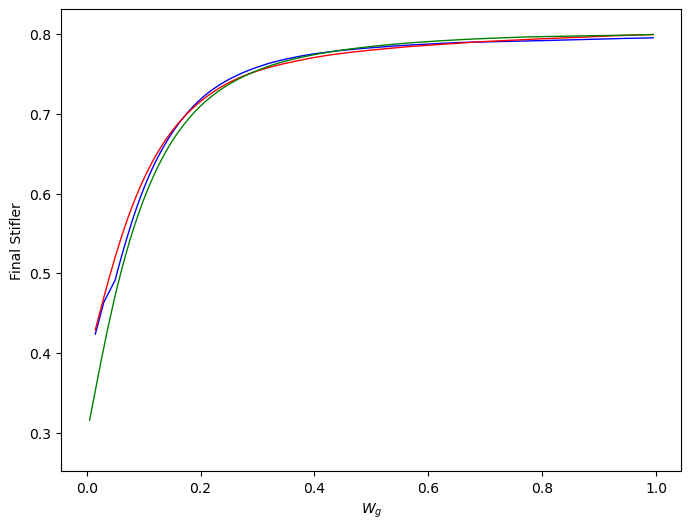} & \includegraphics[width=0.5\textwidth]{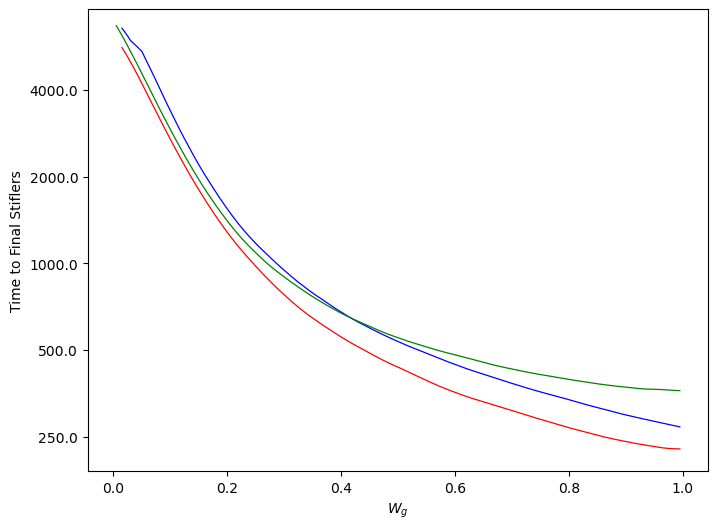}
  \end{tabular}
    \caption{max spreader proportion(top left), time to max spreader(top right), final stifler proportion (bottom left), and time to final stifler (bottom right) in the community (green), random(red) and small-world (blue) networks.}
    \label{fig:combined rumour metrics}
\end{figure}

Figure  \ref{fig:combined rumour metrics} shows the change of the four rumour spreading metrics as a function of the within group connectivity, $w_g$, in the different networks: random (red), community (green) and small-world (blue)

\subsection{Comparing Rumour Spreading Measures}

\onecolumngrid

\twocolumngrid

\paragraph{Max Spreaders (top left)}
The random and small world network both reach an equilibrium where the maximum number of spreaders reaches a maximum value. The community based network is markedly different in that initially the maximum number of spreaders increases but rather than reaching an equilibrium the maximum number of spreaders decreases as a spreader is more likely to meet another spreader or stifler within their community as the intra community links increase. The kink at low $w_g$ for the small-world network is due to converting $w_g$ to a discrete value for $p$, the number of neighbours in the rewiring process. 

\paragraph{Time to Max Spreaders (top right)}
All three networks exhibit a decreasing trend which is intuitive since higher overall connectivity leads to a faster propagation of information on the network. The small world network consistently takes longer to reach equilibrium than the random network because there are relatively fewer long range connections with which information can be spread. The community based network exhibits a switching behaviour in that at low values of $w_g$ it resembles a random network and like a small world network for high $w_g$.

\paragraph{Final Stiflers (bottom left)}
The final number of stiflers  reflects the proportion of population who has once heard the rumour.  All three networks display a consistent pattern which suggests that the final proportion of stiflers is largely independent of the network type under the conditions modelled.

\paragraph{Time to Final Stiflers (bottom right)}
The time when the number of stiflers reaches an equilibrium decreases as $w_g$ increases as more connections in the network (regardless of whether the connections are within or between group) results in a faster spread of the rumour and thus the number of stiflers reaches a maximum. Similar to the time to reach a maximum number of spreaders, the community based network switches from behaving like a random network to a small-world one.

These measures demonstrate the influence of network structure on information dissemination: while structural differences have only a small difference in the ultimate coverage, they have a pronounced impact on the pathway to saturation, shaping the speed and peak intensity of spreading. To explore why this may be the case let us investigate some important structural metrics: the clustering coefficient, modularity and local efficiency along the same parameter space used for the spreading dynamics which is $w_g$ from $0$ to $1$ with $b_g$ fixed at $0.05$, where there are pronounced differences between the 3 networks (fig \ref{fig:clust_coeffs}, \ref{fig:modularity}, \ref{fig:local_efficiency}).

\subsection{Comparing Network Structure Measures}

\begin{figure}[ht!]
\centering
\subfloat[Clustering coefficient]{\label{a}\includegraphics[width=.47\linewidth]{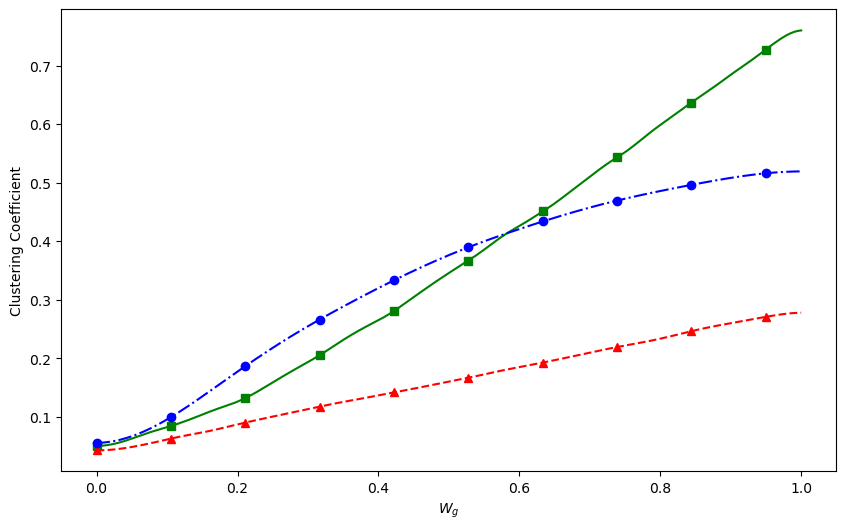}}\hfill
\subfloat[Modularity]{\label{b}\includegraphics[width=.47\linewidth]{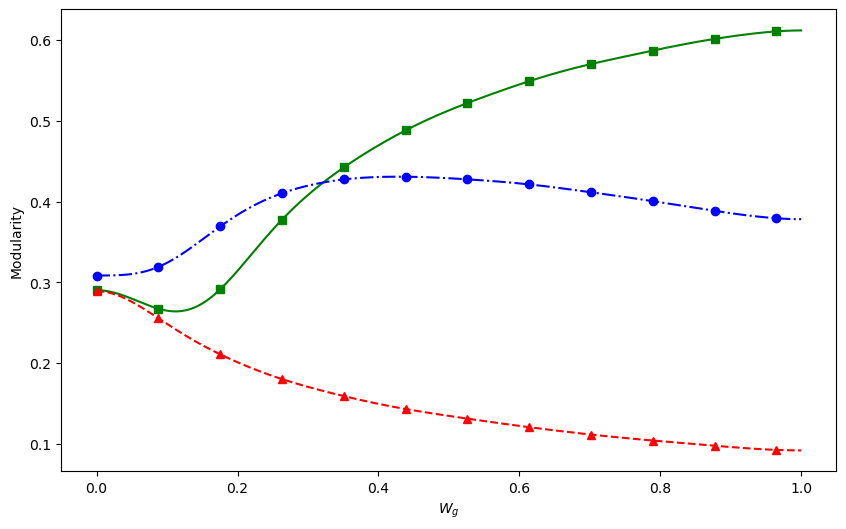}}\par 
\subfloat[Local efficiency]{\label{c}\includegraphics[width=.47\linewidth]{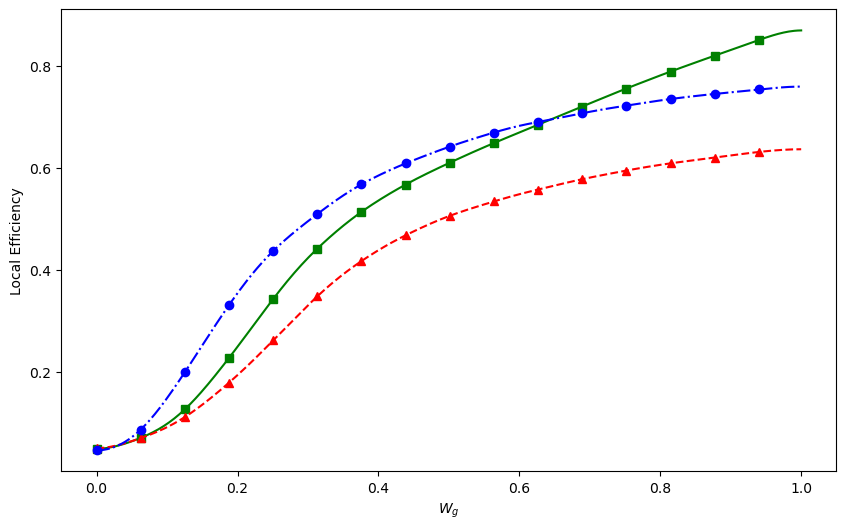}}
\caption{Network measures for random (red), small-world (blue) and community based (green) networks}
\label{fig:measures-wg}
\end{figure}

Figure \ref{fig:measures-wg} present the clustering coefficient, modularity and local efficiency of the networks. These correspond to a cross-sections along $b_g=0.05$ of figures \ref{fig:clust_coeffs},\ref{fig:modularity} and \ref{fig:local_efficiency} respectively. Consistently, the random network has the lowest clustering coefficient, modularity and local efficiencies, while the small-world has the highest measures for low $w_g$, whereas the community based network has the highest for larger $w_g$.

The transmission speed and rumour intensity are negatively correlated with the clustering level. This can be attributed to the stifling mechanism as shown in equation \ref{eqn:rumour_model}. Unlike an epidemic model where the number of infected individuals recover spontaneously, spreaders in the rumour spreading model are inactivated only as a result of contacting other spreaders or stiflers. In this case, the structural difference would lead to a difference of spreaders' stifling efficiency across different types of networks even under unified degree centrality.  In other words, a closely connected local cluster potentially increases the likelihood for a spreader to encounter other spreaders or stifler, thereby reducing its lifespan, which results in a lower rumour intensity in both community and small-world networks compared to the random network. 

\begin{figure}[ht!]
    \centering
    \includegraphics[width=0.95\linewidth]{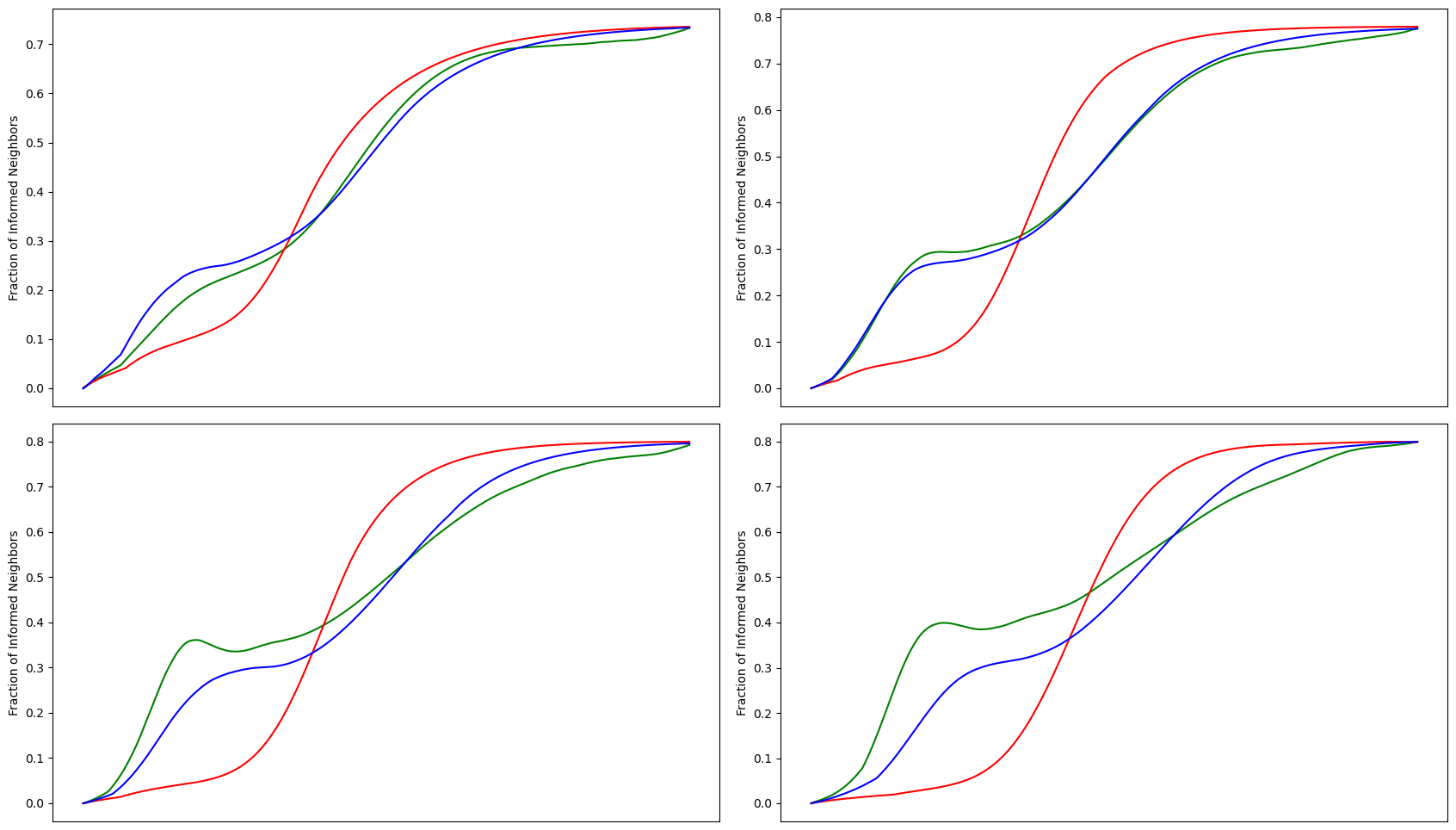}
    \caption{The fractions of informed neighbours for $w_g=0.25$ (top left), $w_g=0.50$ (top right) $w_g=0.75$ (bottom left) and $w_g=1.00$ (bottom right) for random (red), small-world (blue) and community based (green) networks}
    \label{fig:informed 1.00}
\end{figure}

To demonstrate this local saturation effect, we examine how the average proportion of each spreader's informed neighbours varies with \(w_g\), Fig \ref{fig:informed 1.00}. The horizontal axis represents the time and runs until the spreading process has reached its equilibrium. All three networks converge to a similar final level of informed individuals, though their paths differ. We have already seen that the final fraction of informed individuals is independent of the underlying network structure as shown in figure \ref{fig:combined rumour metrics}. 

In the early phase of spreading, community and small-world networks show substantially higher proportions of informed neighbours compared to the random network for each plot in Fig \ref{fig:informed 1.00}, since clustered networks incur a local saturation effect on spreaders. 
The degree of clustering is a critical factor influencing the spreading dynamics across the networks and higher clustering coefficient leads to a higher ratio of spreaders' informed neighbour, which result in shorter lifespan. We find that the speed and intensity of rumour spreading is negatively correlated to the clustering coefficient of the network. In the low $b_g$ high $w_g$ regimen where community networks presents the most pronounced structural characteristics, rumours spread slower than in random or small world networks and with the least intensity.

\subsection{Network Measures for varying connectiveness}
In the preceding analysis, we characterised the community-based network as a collection of sub-groups, with $w_g$ and $b_g$ regulating intra-community and inter-community connections, respectively and compared its structural properties against a random and a small-world network. We see that the community network exhibits a larger clustering coefficient only in the high $w_g$-low $b_g$ regime relative to the reference networks. Additionally, we observe that in this critical regime, rumour propagation is slower than the other networks and with the least intensity. We now investigate whether the observed trends under a fixed transition rate persist across different spreading-stifling ratio. Here, we expand the experimental dimension by introducing spreading-stifling ratio, $\beta$ as a variable for the stability test of robustness in the previous findings. 

 \begin{figure}[ht!]
    \centering
    \includegraphics[width=1\linewidth]{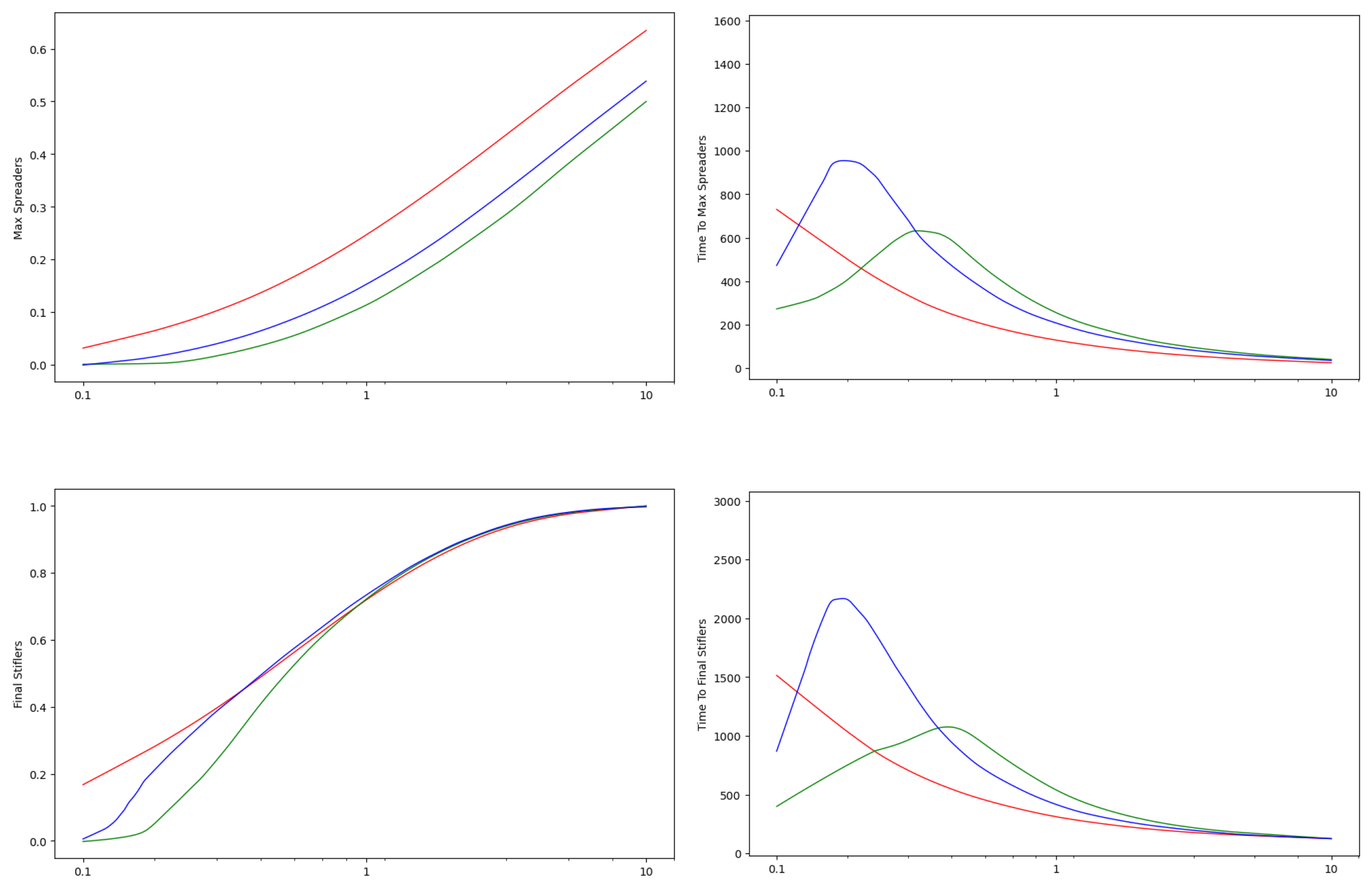}
    \caption{Max spreader proportion (top left), time to max spreaders (top right), final stifler (bottom left) and time to final stifler (bottom right) for community (green), random (red) and small-world (blue) networks. $W_g=0.8,B_g=0.005$, $\beta=(0.1,10)$}
    \label{fig:beta combined}
\end{figure}

We plot the four rumour metrics (max spreader proportion (top left), time to max spreader (top right), final stifler ratio (bottom left) and time to equilibrium (bottom right)) against ratio of spreading and stifling rates, $\beta$, from $0.1$ to $10$, Fig \ref{fig:beta combined}. At $\beta = 1$, it is clear that community network (green) has the lowest max spreader proportion and longest time to both max spreader and time to final stifler; random network (red) consistently has  the highest rumour intensity and spreading speed. Also, the final coverage of the rumour over the three networks are consistent. 
These relative observations are expected according to figure \ref{fig:combined rumour metrics}. However, these trends do not always hold across the $\beta$ domain in every metric. For instance, in both the \textit{time to max spreader} and  \textit{time to final stifler} metrics, the differences among the three networks diminish as $\beta$ increases, eventually converging at high $\beta$ values. However, at low $\beta$, the community and small-world networks exhibit a distinct non-monotonic trend. Initially, both networks start at lower values compared to the random network, but then display an increasing pattern, whereas the random network follows a strictly decreasing trajectory.

Additionally, despite the overall converging tendency as $\beta$ increases, the final stifler proportion demonstrates notable differences across the three networks in the lower $\beta$ regime, which is inconsistent with their relative behaviour observed at $\beta=1$. Notably, the community network exhibits an apparent threshold behaviour in the final stifler metric at low $\beta$, which can be attributed to the local saturation effect. Specifically, in networks with high clustering, rumour propagation is constrained within localised communities, leading to rapid stifling before the rumour can diffuse to other clusters. This mechanism explains the observed non-monotonic trajectories in time to max spreader and time to final stifler for the community and small-world networks.

The underlying process can be interpreted as follows: for a given spreader, the presence of a high fraction of informed neighbours within a densely clustered local community increases the likelihood of transitioning into the stifler state before the rumour can escape to other communities. This effect is particularly pronounced in the low $\beta$ regime, where the spreading rate is significantly lower than the stifling rate. As a result, the overall lifespan of the rumour spreading process is compressed in this regime. If the local saturation effect were absent, the three networks would have exhibited a relative positioning in the low $\beta$ regime that is consistent with their behaviour in the mid-range $\beta$ regime.

Furthermore, the reduction in time to max spreader and time to final stifler is more pronounced in the community network than in the small-world network. This can be attributed to the higher degree of clustering in the community network. That is to say, the stronger local clustering in the community network results in a greater inhibition of spreading, reinforcing the role of local saturation as a fundamental structural constraint on rumour propagation dynamics.

 \begin{figure}[ht!]
    \centering
    \includegraphics[width=1\linewidth]{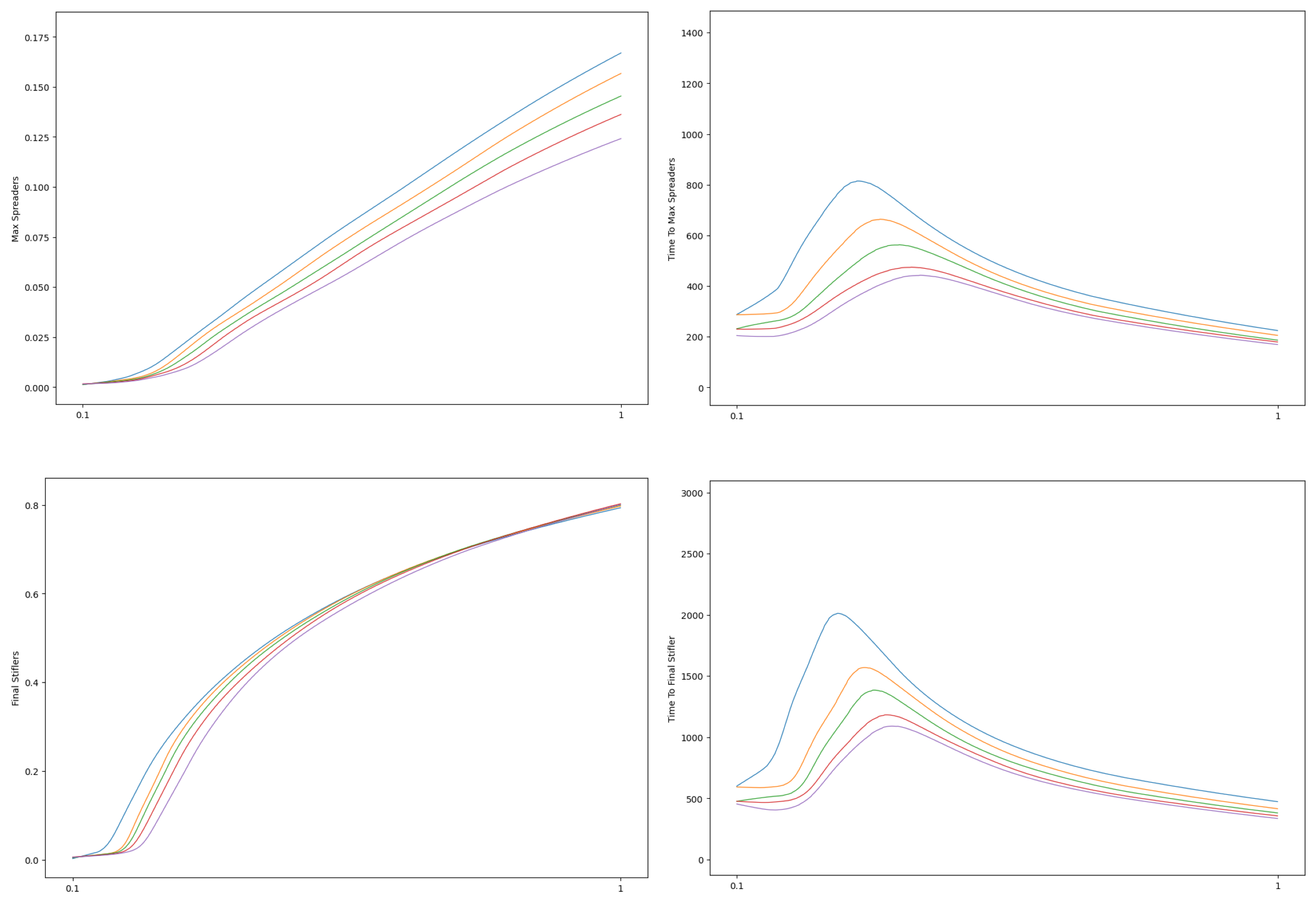}
    \caption{Max spreader proportion(top left), time to max spreader(top right), final stifler(bottom left) and time to final stifler(bottom right) in community network against $\beta$ from $0$ to $1$, with $B_g=0.005$,$W_g=0.6$(blue),$0.7$(orange),$0.8$(green),$0.9$(red), 1.0(purple)}
    \label{fig:community threshold}
\end{figure}

Figure \ref{fig:community threshold} presents the 4 rumour metrics—maximum spreader proportion, time to maximum spreaders, final stifler proportion, and time to final stiflers—as functions of the transmission-to-stifling rate ratio ($\beta$), while varying \(w_g\) ($0.6$ (blue), $0.7$ (orange), $0.8$ (green), $0.9$ (red), 1.0 (purple)). The analysis is conducted solely on the community network, as previous results have demonstrated that threshold behaviour is most pronounced in this network type.

\section{Conclusion}

A simple spreading process like the progress of a rumour through a population behaves differently on different network topologies. 
In this manuscript we have created a community based network to model the spread of information within an idealised group of institutions. Our community based network is based on two parameters which can be related to the node connection probabilities of the small world and random networks that are commonly used to model similar networks. We can see that the community based network has similar properties to both random and small world networks such as the Watts-Strogatz network depending on the location in parameter space. However there are important differences between these networks, particularly when the connectivity between groups is low. This leads to significant differences in how information spreads, particularly when the spread of information within institutions is much greater than between institutions.

By observing the spread of information such as rumours, it may be possible to infer important information about the structure of the underlying network, in particular, when the connections between institutions is low. This can have an important impact on controlling the spread of confidential information and potentially identifying leaks.

\nocite{*}

\bibliography{Rumours_on_CBN}

@PREAMBLE{
 "\providecommand{\noopsort}[1]{}" 
 # "\providecommand{\singleletter}[1]{#1}%" 
}

@article{WattsStrogatz1998,
  title={Collective dynamics of `small-world' networks},
  author={Watts, Duncan J and Strogatz, Steven H},
  journal={nature},
  volume={393},
  number={6684},
  pages={440--442},
  year={1998},
  publisher={Nature Publishing Group}
}

@article{zhang_influence_2018,
	author = {Zhang, Junhuan},
	title = {Influence of individual rationality on continuous double auction markets with networked traders},
	journal = {Physica A: Statistical Mechanics and its Applications},
	volume = {495},
	pages = {353--392},
	year = {2018},
	issn = {0378-4371},
	doi = {https://doi.org/10.1016/j.physa.2017.12.098},
}

@article{d4387031-784a-3051-a366-6206a41ccb7c,
 ISSN = {00219002},
 URL = {http://www.jstor.org/stable/3213787},
 author = {Aidan Sudbury},
 journal = {Journal of Applied Probability},
 number = {2},
 pages = {443--446},
 publisher = {Applied Probability Trust},
 title = {The Proportion of the Population Never Hearing a Rumour},
 urldate = {2024-05-27},
 volume = {22},
 year = {1985}
}

@article{nekovee_theory_2007,
	title = {Theory of rumour spreading in complex social networks},
	volume = {374},
	issn = {0378-4371},
	url = {https://www.sciencedirect.com/science/article/pii/S0378437106008090},
	doi = {https://doi.org/10.1016/j.physa.2006.07.017},
	number = {1},
	journal = {Physica A: Statistical Mechanics and its Applications},
	author = {Nekovee, M. and Moreno, Y. and Bianconi, G. and Marsili, M.},
	year = {2007},
	pages = {457--470},
}

@article{10.1093/imamat/1.1.42,
    author = {DALEY, D. J. and KENDALL, D. G.},
    title = "{Stochastic Rumours}",
    journal = {IMA Journal of Applied Mathematics},
    volume = {1},
    number = {1},
    pages = {42-55},
    year = {1965},
    month = {03},
    issn = {0272-4960},
    doi = {10.1093/imamat/1.1.42},
    url = {https://doi.org/10.1093/imamat/1.1.42},
    eprint = {https://academic.oup.com/imamat/article-pdf/1/1/42/2460250/1-1-42.pdf}
}

@article{zhao_self-organizing_2013,
	author = {Zhao, Haijie and Zhou, Jie and Zhang, Anghui and Su, Guifeng and Zhang, Yi},
	title = {Self-organizing {Ising} model of artificial financial markets with small-world network topology},
	journal = {Europhysics Letters},
	volume = {101},
	number = {1},
	pages = {18001},
	year = {2013},
	doi = {10.1209/0295-5075/101/18001},
	note = {Publisher: EDP Sciences, IOP Publishing and Società Italiana di Fisica},
}

@article{Wan2017,
	author = {Wan, C. and Li, T. anf Sun, Z.},
	title = {Global stability of a SEIR rumor spreading model with demographics on scale-free networks.},
	journal = {Adv Differ Equ},
	year = {2017},
	volume = {253},
	doi = {10.1186/s13662-017-1315-y},
}

@InProceedings{Brisson2018,
author="Brisson, Laurent
and Collard, Philippe
and Collard, Martine
and Stattner, Erick",
editor="Cherifi, Chantal
and Cherifi, Hocine
and Karsai, M{\'a}rton
and Musolesi, Mirco",
title="Information Dissemination in Scale-Free Networks: Profusion Versus Scarcity",
booktitle="Complex Networks {\&} Their Applications VI",
year="2018",
publisher="Springer International Publishing",
address="Cham",
pages="909--920",
isbn="978-3-319-72150-7"
}

@article{hu_hybrid_2023,
	title = {A {Hybrid} {Clustered} {SFLA}-{PSO} algorithm for optimizing the timely and real-time rumor refutations in {Online} {Social} {Networks}},
	volume = {212},
	issn = {0957-4174},
	doi = {https://doi.org/10.1016/j.eswa.2022.118638},
	journal = {Expert Systems with Applications},
	author = {Hu, Xi and Xiong, Xin and Wu, You and Shi, Mengji and Wei, Peng and Ma, Chunmei},
	year = {2023},
	pages = {118638},
}

@article{Sun2021,
  title = {An uncertain SIR rumor spreading model},
  volume = {2021},
  ISSN = {1687-1847},
  url = {http://dx.doi.org/10.1186/s13662-021-03386-w},
  DOI = {10.1186/s13662-021-03386-w},
  number = {1},
  journal = {Advances in Difference Equations},
  publisher = {Springer Science and Business Media LLC},
  author = {Sun,  Hang and Sheng,  Yuhong and Cui,  Qing},
  year = {2021},
  month = jun 
}

@article{Yu2024,
  title = {Dynamic modeling and simulation of double-rumor spreaders in online social networks with IS2TR model},
  volume = {113},
  ISSN = {1573-269X},
  url = {http://dx.doi.org/10.1007/s11071-024-09538-3},
  DOI = {10.1007/s11071-024-09538-3},
  number = {5},
  journal = {Nonlinear Dynamics},
  publisher = {Springer Science and Business Media LLC},
  author = {Yu,  Zhenhua and Zi,  Haiyan and Zhang,  Yun and Wu,  Shixing and Cong,  Xuya and Mostafa,  Almetwally M.},
  year = {2024},
  month = apr,
  pages = {4369–4393}
}

@article{wang2021spreading,
  title={Spreading dynamics of a 2sih2r, rumor spreading model in the homogeneous network},
  author={Wang, Yan and Qing, Feng and Chai, Jian-Ping and Ni, Ye-Peng},
  journal={Complexity},
  volume={2021},
  number={1},
  pages={6693334},
  year={2021},
  publisher={Wiley Online Library}
}

@article{al2015qualitative,
  title={Qualitative analysis of a rumor transmission model with incubation mechanism},
  author={Al-Tuwairqi, Salma and Al-Sheikh, Sarah and Al-Amoudi, Reem},
  journal={Open Access Library Journal},
  volume={2},
  number={11},
  pages={1--12},
  year={2015},
  publisher={Scientific Research Publishing}
}

@article{tong2022dynamic,
  title={Dynamic analysis and optimal control of rumor spreading model with recurrence and individual behaviors in heterogeneous networks},
  author={Tong, Xinru and Jiang, Haijun and Chen, Xiangyong and Yu, Shuzhen and Li, Jiarong},
  journal={Entropy},
  volume={24},
  number={4},
  pages={464},
  year={2022},
  publisher={MDPI}
}

@article{liu2024dynamic,
  title={Dynamic analysis and optimum control of a rumor spreading model with multivariate gatekeepers},
  author={Liu, Yanchao and Zhang, Pengzhou and Li, Deyu and Gong, Junpeng},
  journal={AIMS Mathematics},
  volume={9},
  number={11},
  pages={31658--31678},
  year={2024}
}

@article{pastor2001epidemic,
  title={Epidemic spreading in scale-free networks},
  author={Pastor-Satorras, Romualdo and Vespignani, Alessandro},
  journal={Physical review letters},
  volume={86},
  number={14},
  pages={3200},
  year={2001},
  publisher={APS}
}

@book{keeling2008modeling,
  title={Modeling infectious diseases in humans and animals},
  author={Keeling, Matt J and Rohani, Pejman},
  year={2008},
  publisher={Princeton university press}
}

@article{banerjee_economics_1993,
	title = {The {Economics} of {Rumours}},
	volume = {60},
	issn = {0034-6527},
	doi = {10.2307/2298059},
	number = {2},
	journal = {The Review of Economic Studies},
	author = {Banerjee, Abhijit V.},
	month = apr,
	year = {1993},
	note = {\_eprint: https://academic.oup.com/restud/article-pdf/60/2/309/4476574/60-2-309.pdf},
	pages = {309--327},
}

@article{DANG2021377,
author = {Zhongkai Dang and Lixiang Li and Wei Ni and Renping Liu and Haipeng Peng and Yixian Yang},
title = {How does rumor spreading affect people inside and outside an institution},
journal = {Information Sciences},
volume = {574},
pages = {377-393},
year = {2021},
issn = {0020-0255},
doi = {https://doi.org/10.1016/j.ins.2021.05.085},
}

@article{10.2307/2786545,
 author = {Jeffrey Travers and Stanley Milgram},
 title = {An Experimental Study of the Small World Problem},
 journal = {Sociometry},
 volume = {32},
 number = {4},
 pages = {425--443},
 year = {1969}
}

@article{tanimoto_multi-community_2016,
	author = {Tanimoto, Jun},
	title = {A multi-community homogeneous small-world network and its fundamental characteristics},
	journal = {Physica A: Statistical Mechanics and its Applications},
	volume = {460},
	pages = {88--97},
	doi = {https://doi.org/10.1016/j.physa.2016.04.044},
	year = {2016},
}

@article{Newman2002,
  title = {Assortative Mixing in Networks},
  author = {Newman, M. E. J.},
  journal = {Phys. Rev. Lett.},
  volume = {89},
  issue = {20},
  pages = {208701},
  numpages = {4},
  year = {2002},
  month = {Oct},
  publisher = {American Physical Society},
  doi = {10.1103/PhysRevLett.89.208701},
  url = {https://link.aps.org/doi/10.1103/PhysRevLett.89.208701}
}

@article{zhang_modeling_2010,
	author = {Zhang, Junhuan and Wang, Jun},
	title = {Modeling and simulation of the market fluctuations by the finite range contact systems},
	journal = {Simulation Modelling Practice and Theory},
	volume = {18},
	number = {6},
	pages = {910--925},
	year = {2010},
	doi = {https://doi.org/10.1016/j.simpat.2010.02.008},
	issn = {1569-190X},
}

@article{yang_numerical_2015,
	author = {Yang, Ge and Wang, Jun and Fang, Wen},
	title = {Numerical analysis for finite-range multitype stochastic contact financial market dynamic systems},
	journal = {Chaos: An Interdisciplinary Journal of Nonlinear Science},
	volume = {25},
	number = {4},
	month = apr,
	year = {2015},
	issn = {1054-1500},
	doi = {10.1063/1.4917550},
	note = {\_eprint: https://pubs.aip.org/aip/cha/article-pdf/doi/10.1063/1.4917550/13480587/043111\_1\_online.pdf},
}

@article{barrio_modelling_2017,	
    author = {Barrio, Rafael A. and Govezensky, Tzipe and Ruiz-Guti\'{e}rrez, \'{E}lfego and Kaski, Kimmo K.},
	title = {Modelling trading networks and the role of trust},
	journal = {Physica A: Statistical Mechanics and its Applications},
	volume = {471},
	pages = {68--79},
	doi = {https://doi.org/10.1016/j.physa.2016.11.144},
	year = {2017},
}

@article{cho_trading_2021,
	title = {Trading networks of price-taking buyers and sellers},
	volume = {196},
	issn = {0022-0531},
	doi = {https://doi.org/10.1016/j.jet.2021.105290},
	journal = {Journal of Economic Theory},
	author = {Cho, Myeonghwan},
	year = {2021},
	pages = {105290},
}

@article{gligorijevic_structure_2013,
	author = {Gligorijević, Vladimir and Skowron, Marcin and Tadić, Bosiljka},
	title = {Structure and stability of online chat networks built on emotion-carrying links},
	journal = {Physica A: Statistical Mechanics and its Applications},
	volume = {392},
	number = {3},
	year = {2013},
    doi = {https://doi.org/10.1016/j.physa.2012.10.003},
	pages = {538--543},
}

@article{ElBhih2024,
author = {El Bhih, Amine and  Yaagoub, Zakaria and Rachik, Mostafa and  Allali, Karam and Abdeljawad,Thabet},
title = {Controlling the dissemination of rumors and antirumors in social networks: a mathematical modeling and analysis approach},
journal = {Eur. Phys. J. Plus},
volume = {139},
pages = {118},
year = {2024},
doi = {https://doi.org/10.1140/epjp/s13360-023-04844-y},
}

@article{Escalante2020,
author = {Escalante, Ren\'{e}  and Odehnal, Marco},
title = {A deterministic mathematical model for the spread of two rumors.},
journal = {Afrika Matematika },
volume = {31},
pages = {315-331},
year = {2020},
doi = {https://doi.org/10.1007/s13370-019-00726-8},
}

@book{MakiThompson1973,
  title={Mathematical Models and Applications: With Emphasis on the Social, Life, and Management Sciences},
  author={Maki, D.P. and Thompson, M.},
  isbn={9780135616703},
  lccn={73004471},
  url={https://books.google.co.uk/books?id=xtxLAAAAMAAJ},
  year={1973},
  publisher={Prentice-Hall}
}

@article{ErdosRenyi1959,
author = {Erd\"{o}s, P.; R\'{e}nyi, A.},
title = {On Random Graphs. I},
journal = {Publicationes Mathematicae.},
volume = {6},
number = {3-4},
pages = {290-297},
year = {1959},
}

@article{Rapoport1948,
author = {Rapoport, A.},
title = {Spread of information through a population with socio-structural bias: I. Assumption of transitivity.},
journal = {Bulletin of Mathematical Biophysics},
volume = {15},
pages = {523-533},
year = {1953},
doi = {https://doi.org/10.1007/BF02476440},
}

@article{DaleyKendall1965,
    author = {Daley, D. J. and Kendall, D. G.},
    title = {Stochastic Rumours},
    journal = {IMA Journal of Applied Mathematics},
    volume = {1},
    number = {1},
    pages = {42-55},
    year = {1965},
    month = {03},
    issn = {0272-4960},
    doi = {10.1093/imamat/1.1.42},
}

@article{Zanette2001,
  title = {Dynamics of rumor propagation on small-world networks},
  author = {Zanette, Dami\'an H.},
  journal = {Phys. Rev. E},
  volume = {65},
  issue = {4},
  pages = {041908},
  numpages = {9},
  year = {2002},
  month = {Mar},
  publisher = {American Physical Society},
  doi = {10.1103/PhysRevE.65.041908},
  url = {https://link.aps.org/doi/10.1103/PhysRevE.65.041908}
}

@article{KermackMcKendrick1927,
author = {Kermack, W. O. and McKendrick, A. G.},
title = {A contribution to the mathematical theory of epidemics},
journal = {Proceedings of the Royal Society of London. Series A, Containing Papers of a Mathematical and Physical Character},
volume = {115},
number = {772},
pages = {700-721},
year = {1927},
doi = {https://doi.org/10.1098/rspa.1927.0118},
}

@article{Barabasi1999,
author = {Albert-L\'{a}szl\'{o} Barab\'{a}si  and Réka Albert},
title = {Emergence of Scaling in Random Networks},
journal = {Science},
volume = {286},
number = {5439},
pages = {509-512},
year = {1999},
doi = {10.1126/science.286.5439.509}
}

\end{document}